\documentstyle[12pt]{article}
\newcommand{\z}{&&\hspace*{-1cm}}
\begin{document}

\setcounter{page}{0}
\thispagestyle{empty}

{}~\vspace{2cm}

\begin{center} {\Large {\bf ABOUT THE BEHAVIOUR OF QCD POMERON IN
      DEEP INELASTIC SCATTERING}} \\
 \vspace{1.5cm}
 {\large
  A.V.Kotikov\footnote{
On leave of absence from Particle Physics Laboratory, JINR, Dubna, Russia.\\
e-mail: KOTIKOV@LAPPHP0.IN2P3.FR, KOTIKOV@SUNSE.JINR.DUBNA.SU} \\
 \vspace {0.5cm}
 {\em Laboratoire Physique Theorique ENSLAPP\\
LAPP, B.P. 110, F-74941, Annecy-le-Vieux Cedex, France}}
\end{center}

\vspace{1.5cm}\noindent
\begin{center} {\bf Abstract} \end{center}

 The paper presents the QCD description of the small $x$ behaviour
 of parton distribution functions in the leading twist of Wilson
 operator product expansion. The smooth transition
 between the cases of the soft and hard Pomerons is obtained. The
 results are in qualitative agreement with deep inelastic experimental
 data.

PACS number(s): 13.60.Hb, 12.38.Bx, 13.15.Dk

\newpage
\pagestyle{plain}

\section{Introduction} \indent

The recent measurements of the deep-inelastic (DIS) structure function
(SF) $F_2$ by the $H1$ \cite{1} and $ZEUS$ \cite{2} collaborations
open a new kinematical range to study proton structure. The new
$HERA$ data show the strong increase of $F_2$ with decreasing
$x$, that contradicts to experimental data of the $NMC$ \cite{3} and
$E665$ collaboration \cite{4} at lower $Q^2$ ($Q^2 \sim 1 GeV^2$,
which are quite flat at $x \sim 10^{-2}$. These lower $Q^2$ data are
 in the good agreement with
the standard Pomeron or with the Donnachie-Landshoff picture
\cite{4.5} where the
Pomeron intercept $\alpha_p = 1.08$, is very close to standard
one $\alpha_p = 1$.
However, the $HERA$ data extracted at the larger $Q^2$ ($Q^2 > 10
GeV^2$) are fitted very well (see \cite{19,17}) by parametrizations of
parton distributions (PD) contained the supercritical (or Lipatov) Pomeron.
The interpritation of the fast changing
of the intercept in
the $Q^2$ region between $Q^2=1GeV^2$ and $Q^2=10GeV^2$ (see
\cite{5}) is yet absent. There are
arguments (see \cite{6}) in favour of one intercept or a
superposition of two different Pomeron trajectories, one having an
intercept of $1.08$ and the one of $1.5$ (see discussions of this
problem in \cite{5}).

The aim of this letter is a possible ``solution'' of this problem in
the framework of
Dokshitzer-Gribov-Lipatov-Altarelli-Parisi (DGLAP) equation
\cite{6.5}.
 We will seek the
``solution''\footnote{We use
  the termin ``solution'' because we will work in the leading twist
  approximation in the range of $Q^2$: $Q^2>1GeV^2$, where the higher
  twist terms may give the sizeable contribution.
  Moreover, our ``solution'' is the Regge asymptotic
  (\ref{1}) with
  unknown parameters rather then the solution of DGLAP equation. The
  parameters are found from the agreement of the r.h.s. and l.h.s. of
  the equation.}
 of DGLAP
equation imposing the Regge-type behaviour
 (we use the parton distributions (PD)
multiplied by $x$ and neglect the nonsinglet quark distribution at
small $x$):
 \begin{eqnarray}
f_a (x,Q^2) \sim x^{-\delta} \tilde
f_a(x,Q^2), ~~~~(a=q,g)~~(\alpha_p \equiv 1+\delta ) \label{1}
 \end{eqnarray}
 where $\tilde f_a(x,Q^2)$ is nonsingular at $x \to 0$ and $\tilde
f_a(x,Q^2) \sim (1-x)^{\nu}$ at $x \to 1$\footnote{Consideration of
  the more complicate behaviour in the form $x^{-\delta}(ln(1/x))^b
  I_{2g}(\sqrt{\phi ln(1/x)})$ is given in \cite{9a}
}.
Of course, we understand that the Regge behaviour (\ref{1}) contradicts
to the well-known double-logarithmical solution:
$~ \sim \exp{\sqrt{\phi (Q^2) ln(1/x)}}$, where
$\phi (Q^2)$ is a known $Q^2$-dependent function\footnote{More
  correctly, $\phi$ is $Q^2$-dependent for the solution of DGLAP
equations with the  boundary condition: $f_a(x,Q^2_0)= Const$ at
  $x \to 0$. In the case of the boundary condition: $f_a(x,Q^2_0) \sim
  \exp{\sqrt{ ln(1/x)}}$, $\phi$ is lost (see \cite{9a}) its
  $Q^2$-dependence}
. However, we think that it
is not possible to glue the both Regge-type behaviour:
Donnachie-Landshoff picture at low $Q^2$ and Lipatov Pomeron at large
$Q^2$, having the double-logarithmical (non Regge-type) solution at
immediate $Q^2$.

In the case of the large $\delta$ values (i.e. $x^{-\delta} \gg 1$)
 similar investigations were already done
 (see
\cite{12}-\cite{14})
and the results are well known (see \cite{12} for the first two
orders of the perturbation theory, \cite{14} for the first three
orders and \cite{14a} containing a resummation of all orders,
respectively):
 \begin{eqnarray}
 \frac{f_a(x,t)}{f_a(x,t_0)}~=~ \frac{M_a(1+\delta,t)}{M_a(1+\delta,t_0)},
 \label{6}
 \end{eqnarray}
 where $t=ln(Q^2/\Lambda ^2),~
t_0=t(Q^2=Q^2_0)$ and $M_a(1+\delta,t)$ is the analytical continuation
of the PD moments
 $M_a(n,t) = \int^1_0 dx x^{n-2} f_a(x,t)$ to the noninteger
value ``$n=1+\delta$''.
Note that recently the fit of $HERA$ data was
done in \cite{17} with the formula for PD $f_q(x,t)$ very close
\footnote{The used  formula (Eq.(2) from \cite{17}) coincides with
  (\ref{6}) in the leading order (LO) approximation, if we save
only $f_g(x,Q^2)$ in the
  r.h.s. of (\ref{2}) (or put $\gamma_{qq}=0$ and $\gamma_{qg}=0$
  formally). Eq.(\ref{6}) and Eq.(\ref{2}) from \cite{17} have
    some differences in the next-to-leading order (NLO), which are not
very important because they are
    corrections to the $\alpha$-correction.} to
(\ref{6}) and a very good agreement (the $\chi^2$ per degree of freedom
is $0.85$) is found for $\delta = 0.40 \pm 0.03$.
There are also
fits \cite{15a} of the another group using equations which are
similar to (\ref{6}) in the LO approximation.

  In this article we expand
these results to the range where $\delta \sim 0$ (and $Q^2$ is not
large) following to the observed earlier (see \cite{12,13} and Appendix)
method\footnote{The
method is based on the earlier results \cite{10a}}
 to replace the Mellin convolution by a simple product.

Consider DGLAP equations and apply the method from \cite{13} to the
Mellin convolution in its r.h.s. (in contrast with standard case, we
use below $\alpha(Q^2)=\alpha_s(Q^2)/(4\pi)$):
 \begin{eqnarray}
\z \frac{d}{dt}f_a (x,t)~=~- \frac{1}{2} \sum_{i=a,b}
\hat \gamma_{ai}(\alpha,x) \otimes f_i(x,t)~~~(a,b)=(q,g) \nonumber \\
\z =~- \frac{1}{2} \sum_{i=a,b}
\tilde \gamma_{ai}(\alpha,1+\delta)f_i(x,t)~+~O(x^{1-\delta})~~~
\Bigl( \gamma_{ab}(\alpha,n)=\alpha \gamma_{ab}^{(0)}(n)+\alpha^2
\gamma_{ab}^{(1)}(n)+...\Bigr)
, \label{2}
 \end{eqnarray}
where $t=ln(Q^2/\Lambda ^2)$. The $\hat \gamma_{ab}(\alpha,x)$ are the
spliting functions corresponding to the anomalous dimensions (AD)
$\gamma_{ab}(\alpha,n) = \int^1_0 dx x^{n-2} \hat \gamma_{ab}(\alpha,x)$. Here
the functions $\gamma_{ab}(\alpha,1+\delta)$
are the AD $\gamma_{ab}(\alpha,n)$ expanded from the
integer argument ``$n$'' to the noninteger one ``$1+\delta$''. The
functions $\tilde \gamma_{ab}(\alpha,1+\delta)$ (marked below as AD, too)
can be obtained from
the functions $\gamma_{ab}(\alpha,1+\delta)$  replacing the term $1/\delta$
by the one $1/\tilde \delta$:
 \begin{eqnarray}
\frac{1}{\delta} \to  \frac{1}{\tilde \delta}~=~\frac{1}{\delta}
\Bigl( 1 - \varphi(\nu,\delta)x^{\delta} \Bigr)
 \label{3}
 \end{eqnarray}
This replacement (\ref{3}) appeares
very naturally in the consideration of the Mellin convolution at $x \to
0$ (see \cite{13} and Appendix) and
preserves the smooth and nonsingular transition to
the case $\delta =0$, where
 \begin{eqnarray}
 \frac{1}{\tilde \delta}~=~ln\frac{1}{x} - \varrho(\nu)
 \label{4}
 \end{eqnarray}

The concrete form of the functions $\varphi(\nu,\delta)$ and
$\varrho(\nu)$ depends strongly on the type of the behaviour of the
PD $f_a(x,Q^2)$ at $x \to 0$ and in the case of the Regge regime (\ref{1})
they are (see Appendix
):
 \begin{eqnarray}
\varphi(\nu,\delta)~=~  \frac{\Gamma(\nu +1)\Gamma(1-\delta)}{\Gamma(\nu
  +1-\delta)}~~ \mbox{ and }~~ \varrho(\nu)~=~\Psi(\nu+1)-\Psi(1),
 \label{5}
 \end{eqnarray}
where $\Gamma(\nu+1)$ and $\Psi(\nu+1)$ are the Euler $\Gamma$- and
$\Psi$-functions, respectively. As it can be seen, there is
a correlation with the PD behaviour at large $x$.

If $\delta$ is not small (i.e. $x^{-\delta}>>1$), we can replace
$1/ \tilde \delta$ by $1/\delta$ in the r.h.s. of Eq.(\ref{2}) and
obtain its solution in the form (\ref{6}). The case of small $\delta $
values will be considered lower.


The new points in our investigations are as follows. Note that the
$Q^2$-evolution of $M_a(1+\delta,t)$ contains  the two: ``+'' and
``$-$'' components, i.e. $M_a(1+\delta,t)=
\sum_{i= \pm} M_a^i(1+\delta,t)$,
and in principle each component evolves separately and may have
 independent (and not equal) intercept.

\section{Leading order} \indent

Consider DGLAP equations for the ``+'' and ``$-$'' parts
(hereafter $s=ln(lnt/lnt_0)$):
  \begin{eqnarray}
 \frac{d}{ds} f_a^{\pm}(x,t)~=~- \frac{1}{2\beta_0}
\tilde
\gamma_{\pm}(\alpha,1+\delta_{\pm})f_a^{\pm}(x,t)~+~O(x^{1-\delta}),
 \label{7} \end{eqnarray}
where
$$\gamma_{\pm}~=~ \frac{1}{2}
\biggl[
\Bigl(\gamma_{gg}+\gamma_{qq} \Bigr)~\pm ~
\sqrt{ {\Bigl( \gamma_{gg}- \gamma_{qq} \Bigr)}^2
  ~+~4\gamma_{qg}\gamma_{gq}}
\biggr]$$
are the AD of the ``$\pm $'' components (see, for example, \cite{18})

The AD $\tilde \gamma_{-}(\alpha,1+\delta_-)$ does not
contain the singular term (see \cite{12,14} and below) and $f_a^-(x,t)$
have the solution in the form:
 \begin{eqnarray}
 \frac{f_a^-(x,t)}{f_a^-(x,t_0)}~=~e^{-d_-(1+\delta_-)s}, \mbox{ where }
d_{\pm}=\frac{\gamma_{\pm}(1+\delta_{\pm})}{2\beta_0}
 \label{8}
 \end{eqnarray}

The AD $\tilde \gamma_{+}(\alpha,1+\delta_+)$
contains the singular term  and $f_a^+(x,t)$  have the solution similar
(\ref{8})
only for $x^{-\delta_+}>>1$:
 \begin{eqnarray}
 \frac{f_a^+(x,t)}{f_a^+(x,t_0)}~=~e^{-d_+(1+\delta_+)s}, \mbox{ if }
 x^{-\delta_+}>>1
 \label{9}
 \end{eqnarray}

 Both intercepts $1+\delta_+$ and $1+\delta_-$ are
unknown  and should be found, in principle, from the analysis of the
experimental
data. However there is the another way. From the small $Q^2$ (and
small $x$) data of the $NMC$ \cite{3} and $E665$ collaboration \cite{4} we
can conclude that the SF $F_2$ and hence the PD $f_a(x,Q^2)$ have
flat asymptotics for $x \to 0$ and $Q^2 \sim (1\div2)GeV^2$. Thus we
know that the values of $\delta_+$ and $\delta_-$ is approximately
zero at $Q^2 \sim 1GeV^2$.

 Consider  Eqs.(\ref{7}) with
$\delta_{\pm}=0$ and with the boundary condition $f_a(x,Q^2_0)=A_a$ at
$Q^2_0=1GeV^2$. For the ``$-$'' component we already have the solution:
 Eq.(\ref{8}) with $\delta_-=0$ and $d_-(1)=16f/(27\beta_0$), where
$f$ is the number of the active quarks and $\beta_i$ are the
coefficients in the $\alpha$-expansion of QCD $\beta$-function.
For its ``+'' component
Eq.(\ref{7}) can be rewritten in the form (hereafter the index
$1+\delta $ will be omitted in the case $\delta \to 0$):
  \begin{eqnarray} ln(\frac{1}{x})\frac{d}{ds}\delta_+(s)~+~
 \frac{d}{ds} ln(A_a^+) ~=~- \frac{1}{2\beta_0}
\biggr[ \hat
\gamma_{+}
\Bigl( ln(\frac{1}{x}) -\varrho(\nu)  \Bigr) ~+~ \overline \gamma_+
\biggl]
 \label{10} \end{eqnarray}
where
$\hat\gamma_{+}$ and $\overline \gamma_+$ are the coefficients of the
singular and regular parts at $\delta \to 0$ of AD
$\gamma_+(1+\delta)$:
$$ \gamma_+(1+\delta)~=~\hat\gamma_+ \frac{1}{\delta} ~+~
\overline\gamma_+,~~~\hat\gamma_+=-24,~\overline\gamma_+=22+
\frac{4f}{27}$$

The solution of Eq.({10}) is
 \begin{eqnarray}
 f_a^+(x,t)~=~A^+_a~x^{\hat d_+s}e^{-\overline d_+s},
 \label{11}
 \end{eqnarray}
where
$$\hat d_+ \equiv \frac{\hat \gamma_+}{2\beta_0} \simeq -
\frac{4}{3},~~
\overline d_+ \equiv \frac{1}{2\beta_0}
\Bigl( \overline \gamma_+  ~-~ \hat \gamma_+ \varrho(\nu)\Bigr)
 \simeq \frac{4}{3} \varrho(\nu) + \frac{101}{81}$$
Hereafter the symbol $\simeq $ marks the case $f=3$.

As it can be seen from (\ref{11}) the flat form $\delta_+=0$ of the
``+''-component of PD is very nonstable from the (perturbative)
viewpoint, because $d(\delta_+)/ds \neq 0$, and for $Q^2 > Q_0^2$ we
have already the nonzero power of $x$ (i.e. Pomeron intercept
$\alpha_p >1$). This is in agreement with the experimental data. Let us
note that the power of x is positive for $Q^2<Q^2_0$ that is in principle
also supported by the $NMC$ \cite{3}
data, but the use of this analysis to $Q^2<1GeV^2$ is open
question.

Thus, we have the DGLAP equation solution for the ``+'' component at $Q^2$
is close to $Q^2_0=1GeV^2$, where Pomeron starts in its movement to the
supercritical (or Lipatov \cite{19.5,19.6}) regime and also for the large
$Q^2$, where Pomeron have
  $Q^2$-independent intercept. In principle, the general
solution of (\ref{7}) should contain the smooth transition between
these pictures but this  solution is absent
\footnote{The form $\exp \Bigl({ -s \tilde
    \gamma_+(1+\delta)/(2\beta_0)} \Bigr)$ coincides with both solutions:
  Eq.(\ref{9}) if $x^{\hat d_+} >>1$ and Eq.(\ref{11}) when $\delta
  =0$ but it is not the solution of DGLAP equation.}. We
introduce  some ``critical'' value of $Q^2$, $Q^2_c$, where the
solution (\ref{9}) is replaced by the solution (\ref{11}). The
exact value of $Q^2_c$ may be obtained from a fit of experimental
data. Thus, we have in the LO of the perturbation theory:
 \begin{eqnarray}
\z f_a(x,t)~=~ f_a^-(x,t)~+~ f_a^+(x,t),~~
f_a^-(x,t)~=~A^-_a~\exp{(- d_-s)} \nonumber  \\
\z f_a^+(x,t)~=~
\left\{
\begin{array}{ll} A^+_a
x^{\hat d_+s}\exp{(-\overline d_+s)}, & \mbox{ if } Q^2 \leq Q^2_c \\
f_a^+(x,t_c)
\exp{\Bigl(-d_+(1+\delta_c)(s-s_c)\Bigr)},
& \mbox{ if } Q^2>Q^2_c
\end{array} \right.
 \label{12}
 \end{eqnarray}
where
 \begin{eqnarray}
\z t_c~=~t(Q^2_c),~~s_c~=~s(Q^2_c),~~A_a^-~=~A_a ~-~ A_a^+ ~~ \mbox{and }
\nonumber  \\
\z
A^+_q~=~(1- \overline \alpha )A_q ~-~ \tilde \alpha A_g,~~
A^+_g~=~ \overline \alpha A_g ~-~ \varepsilon A_q
 \label{13}
 \end{eqnarray}
and the values of the coefficients $\overline \alpha$, $\tilde \alpha$
and $\varepsilon$ may be found, for example, in \cite{18}.

Using the concrete AD values at $\delta =0$ and $f=3$, we have
 \begin{eqnarray} \z
A^+_q~ \simeq ~\frac{1}{27}\frac{4A_q+9A_g}{ln(\frac{1}{x})-\varrho
  (\nu) - \frac{85}{108}}
,~~A^+_g~ \simeq ~A_g~+~\frac{4}{9}A_q ~-~
\frac{4}{243}\frac{9A_g-A_q}{ln(\frac{1}{x})-\varrho
  (\nu) - \frac{85}{108}}
 \label{14}
 \end{eqnarray}

Thus, the value of the ``+''component of the quark PD is suppressed
logarithmically and this
 is in qualitative agreement with the $HERA$ parametrizations of
SF $F_2$ (see \cite{20.5,20})
(in the LO $F_2(x,Q^2)=(2/9)f_q(x,Q^2)$ for $f=3$), where
the magnitude connected
with the factor $x^{-\delta}$ is $5 \div 10 \%$ from the flat (for $x \to
                                        0$) magnitude.

\section{Next-to-leading order} \indent

By analogy with the
previous section
and knowing the NLO
$Q^2$-dependence of PD moments, we obtain the following equations for the NLO
$Q^2$-evolution of the both: ''+'' and ``$-$'' PD components (hereafter
$\tilde s=ln(\alpha(Q^2_0)/\alpha(Q^2)), p=\alpha(Q^2_0)-\alpha(Q^2)$):
 \begin{eqnarray}
\z f_a(x,t)~=~ f_a^-(x,t)~+~ f_a^+(x,t)
,~~f_a^-(x,t)=~\tilde A^-_a~\exp{(- d_-\tilde s -d_{--}^ap)}
 \nonumber  \\
\z f_a^+(x,t)=
\left\{
\begin{array}{ll} \tilde A^+_a
x^{(\hat d_+\tilde s + \hat d_{++}^a p)}\exp{(-\overline d_+\tilde s
  -\overline d_{++}^ap)}, & \mbox{if } Q^2 \leq Q^2_c \\
f_a^+(x,t_c)
\exp{\Bigl(-d_+(1+\delta_c)(\tilde s-\tilde
  s_c)-d_{++}^a(1+\delta_c)(p-p_c) \Bigr) },
& \mbox{if } Q^2>Q^2_c
\end{array} \right.
 \label{15}
 \end{eqnarray}
where
 \begin{eqnarray} \z
\tilde s_c ~=~ \tilde s(Q^2_c),~
p_c~=~p(Q^2_c),~\alpha_0~=~\alpha(Q^2_0)
,~\alpha_c~=~\alpha(Q^2_c) \nonumber \\ \z
\tilde A^{\pm}_a~=~\Bigl(1~-~\alpha_0 K^a_{\pm} \Bigr)
A^{\pm}_a ~+~ \alpha_0 K^a_{\pm} A^{\mp}_a \nonumber \\  \z
d_{++}^a~=~ \hat d_{++}^a
\Bigl(
ln(\frac{1}{x})
- \varrho(\nu) \Bigl)
{}~+~ \overline d_{++}^a, ~~
d^a_{++}~=~ \frac{\gamma_{\pm \pm}}{2\beta_0} ~-~
\frac{\gamma_{\pm} \beta_1}{2\beta^2_0} ~-~ K^a_{\pm}  \nonumber \\
\z\mbox{and }~~ K^q_{\pm}~=~ \frac{\gamma_{\pm \mp}}{2\beta_0 +
  \gamma_{\pm} - \gamma_{\mp}},~~ K^g_{\pm}~=~ K^q_{\pm}
\frac{\gamma_{\pm}-\gamma^{(0)}_{qq}}{\gamma_{\mp}-\gamma^{(0)}_{qq}}
 \label{16}
 \end{eqnarray}

The NLO AD of the ``$\pm$'' components are connected with the NLO AD
$\gamma^{(1)}_{ab}$. The corresponding formulae can be found in
\cite{18}.

Using the concrete values of the LO and  NLO AD at $\delta =0$ and
$f=3$, we obtain the following values for the NLO components from
(\ref{15}),(\ref{16}) (note that we keep only the terms $\sim O(1)$
in the NLO terms)
 \begin{eqnarray} \z
 d^q_{--}~\simeq~ \frac{16}{81} \Big[ 2\zeta (3) + 9 \zeta (2) -
\frac{779}{108}  \Big] \approx 1.97, ~~
d^g_{--}~\simeq~ d^q_{--}~+~ \frac{28}{81} \approx 2.32 \nonumber \\ \z
\hat d^q_{++}~\simeq~ \frac{2800}{81} , ~~
\overline d^q_{++}~\simeq~ 32 \Big[ \zeta (3) + \frac{263}{216}\zeta (2) -
\frac{372607}{69984}  \Big] \approx -67.82    \nonumber \\ \z
\hat d^g_{++}~\simeq~ \frac{1180}{81} , ~~
\overline d^g_{++}~\simeq~ \overline d^q_{++}~+~ \frac{953}{27} -12\zeta
(2) \approx -52.26
 \label{17}
 \end{eqnarray}
and
 \begin{eqnarray} \z
\tilde A^+_q~ \simeq ~\frac{20}{3} \alpha_0
\Bigl[ A_g + \frac{4}{9} A_q \Bigr] ~+~
\frac{1}{27}\frac{4A_q(1-7.68 \alpha_0)+9A_g(1-8.72
  \alpha_0)}{ln(\frac{1}{x})-\varrho
  (\nu) - \frac{85}{108}}  \nonumber \\
\z \tilde A^+_g~ \simeq ~ A_g \Bigl(1-\frac{80}{9}\alpha_0 \Bigr)
 ~+~\frac{4}{9} \biggl[A_q
 ~-~
\frac{1}{27}\frac{9A_g-A_q}{ln(\frac{1}{x})-\varrho
  (\nu) - \frac{85}{108}} \biggr]  \Bigl( 1- \frac{692}{81}\alpha_0)
 \nonumber \\
\z\mbox{and } \tilde A_a^-~=~A_a ~-~ \tilde A_a^+
 \label{18}
 \end{eqnarray}

It is useful to change  in Eqs.(\ref{15})-(\ref{18}) from the quark
PD to the SF $F_2(x,Q^2)$, which is connected in NLO approximation with the PD
in the following way (see \cite{18}):
 \begin{eqnarray}
F_2(x,Q^2)~=~ \Bigl(  1+\alpha(Q^2)B_q(1+\delta) \Bigr) \delta^2_s
f_q(x,Q^2) ~+~ \alpha(Q^2)B_g(1+\delta) \delta^2_s f_g(x,Q^2),
 \label{19}
 \end{eqnarray}
where $\delta ^2_s = \sum_{i=1}^f/f \equiv <e_f^2>$ is the average
charge square of the active quarks: $\delta ^2_s$ = (2/9 and 5/18) for
$f$ = (3 and 4), respectively.
 The NLO corrections lead to the
appearence in the r.h.s. of Eqs.(\ref{15}) of the additional terms
$\Bigl(  1+\alpha B_{\pm} \Bigr)/\Bigl(  1+\alpha_0 B_{\pm} \Bigr)$
and the necessity to transform
$\tilde A^{\pm}_q$ to $C^{\pm} \equiv F_2^{\pm}(x,Q^2)$ into the input
parts. The final results for $F_2(x,Q^2)$ are in the form:
 \begin{eqnarray}
\z F_2(x,t)~=~ F_2^-(x,t)~+~ F_2^+(x,t) \nonumber  \\
\z F_2^-(x,t)~=~ C^-~\exp{(- d_-\tilde s -d_{--}^qp)}
(1+\alpha B^-)/(1+\alpha_0 B^-)
 \nonumber  \\
\z F_2^+(x,t)~=~
\left\{
\begin{array}{ll} C^+
x^{(\hat d_+\tilde s + \hat d_{++}^q p)}\exp{(-\overline d_+\tilde s
  -\overline d_{++}^qp)}(1+\alpha B^+)/(1+\alpha_0 B^+)
, & \mbox{ if } Q^2 \leq Q^2_c \\
F_2^+(x,t_c)
\exp{\Bigl(-d_+(1+\delta_c)(\tilde s-\tilde
  s_c)-d_{++}^q(1+\delta_c)(p-p_c) \Bigr) } &  \\
\biggl(1+
\alpha B^+(1+\delta_c) \biggr)/
\biggl(1+
\alpha_c B^+(1+\delta_c) \biggr),
& \mbox{ if } Q^2>Q^2_c
\end{array} \right.
 \label{20}
 \end{eqnarray}
where
$$ B^{\pm}~=~B_q ~+~ \frac{\gamma_{\pm}}{\gamma^{(0)}_{qg}}B_g,~~
C^{\pm}~=~\tilde A^{\pm}_q (1+\alpha_0 B^{\pm})$$
with the substitution of $A_q$ by $C \equiv F_2(x,Q^2_0)$ into
Eq.(\ref{18}) $\tilde
A^{\pm}_q$ according
 \begin{eqnarray} \z
C~=~ \Bigl(  1+\alpha_0 B_q \Bigr) \delta^2_s
A_q ~+~ \alpha_0 B_g \delta^2_s A_g,
 \label{21}
 \end{eqnarray}

For the gluon PD the situation is more simple: in
Eq.(\ref{18}) it is necessary to replace $A_q$ by $C$
according (\ref{21}).

For the concrete values of the LO and NLO AD at $\delta =0$ and $f=3$,
we have for $Q^2$-evolution of $F_2(x,Q^2)$ and the gluon PD:
 \begin{eqnarray}
\z F_2(x,t)~=~ F_2^-(x,t)~+~ F_2^+(x,t),~~
f_g(x,t)~=~ f_g^-(x,t)~+~ f_g^+(x,t) \nonumber  \\
\z F_2^-(x,t)=~ C^-~\exp{(- \frac{32}{81} \tilde s
  -1.97p)}(1-\frac{8}{9} \alpha )/(1-\frac{8}{9} \alpha_0 )
 \nonumber  \\
\z F_2^+(x,t)=
\left\{
\begin{array}{ll} C^+
x^{(-\frac{4}{3} \tilde s + \frac{2800}{81}
  p)}
\exp{\Bigl(- \frac{4}{3}(\varrho(\nu)+\frac{101}{108}) \tilde s
  +(\frac{2800}{81} \varrho(\nu)+67.82)p \Bigr)}  &  \\
\Bigl(1+6[ln(\frac{1}{x})-\varrho(\nu)-\frac{101}{108}] \alpha \Bigr)/
\Bigl(1+6[ln(\frac{1}{x})-\varrho(\nu)-\frac{101}{108}] \alpha_0 \Bigr)
, & \mbox{if} Q^2 \leq Q^2_c \\
F_2^+(x,t_c)
\exp{\Bigl(-d_+(1+\delta_c)(\tilde s-\tilde
  s_c)-d_{++}^q(1+\delta_c)(p-p_c) \Bigr) }  &  \\
\biggl(1+
\alpha B^+(1+\delta_c) \biggr)/
\biggl(1+
 \alpha_c B^+(1+\delta_c) \biggr),
& \mbox{if} Q^2>Q^2_c
\end{array} \right.
 \label{22}  \\
\z f_g^-(x,t)=~ \tilde A_g^-~\exp{(- \frac{32}{81} \tilde s
  -2.32p)}
 \nonumber  \\
\z f_g^+(x,t)=
\left\{
\begin{array}{ll} \tilde A_g^+
x^{(-\frac{4}{3} \tilde s + \frac{1180}{81}
  p)}\exp{\Bigl(- \frac{4}{3}(\varrho(\nu)+\frac{101}{108}) \tilde s
  +(\frac{1180}{81} \varrho(\nu)+52.26)p \Bigr)}
, & \mbox{if} Q^2 \leq Q^2_c \\
f_g^+(x,t_c)
\exp{\Bigl(-d_+(1+\delta_c)(\tilde s-\tilde
  s_c)-d_{++}^g(1+\delta_c)(p-p_c) \Bigr) }
,& \mbox{if} Q^2>Q^2_c
\end{array} \right.
 \label{23}
 \end{eqnarray}
where
 \begin{eqnarray} \z C^- ~=~ C ~-~ C^+,~~\tilde A_g^-~=~A_g ~-~ \tilde
   A_g^+ ~~~
\mbox{and } \nonumber  \\  \z
C^+~ \simeq ~\frac{2}{27}
\Biggl(  26\alpha_0
\Bigl[ A_g + 2C \Bigr] ~+~
\frac{A_g(1-10.50 \alpha_0)+2C(1-8.55
  \alpha_0)}{ln(\frac{1}{x})-\varrho
  (\nu) - \frac{85}{108}} \Biggr) \nonumber \\
\z \tilde A^+_g~ \simeq ~A_g \Bigl(1-\frac{88}{9}\alpha_0 \Bigr)
{}~+~2C\Bigl(1-\frac{692}{81}\alpha_0 \Bigr) ~-~
\frac{2}{27}\frac{2A_g(1-\frac{674}{81}\alpha_0)-
C(1- \frac{692}{81}\alpha_0)}{ln(\frac{1}{x})-\varrho
  (\nu) - \frac{85}{108}}
 \label{25}
 \end{eqnarray}

Let us give some conclusions following from
Eqs.(\ref{22})-(\ref{25}). It is clearly seen that the NLO
corrections reduce the LO contributions. Indeed, the value of the
supercritical Pomeron intercept,
which increases as $ln(\alpha_0/\alpha)$ in the
LO,  obtains the additional  term $ \sim (\alpha_0 - \alpha)$ with the
large (and opposite in sign to the LO term) numerical coefficient. Note
that this coefficient is different for the quark and gluon PD and  this is
in agreement with the recent $MRS(G)$ fit in \cite{19} and the
data analysis by $ZEUS$ group (see \cite{20}). The intercept of the
gluon PD is larger then the quark PD one (see also \cite{19,20}).
However, the effective reduction of the quark PD is smaller (which is
in agreement with W.-K. Tung analysis in \cite{16a}), because the quark PD
 part increasing at small $x$  obtains the additional ($ \sim
\alpha_0$ but not $ \sim 1/lnx $) term, which is  important at very small
$x$.

Note that there is the fourth quark threshold at $Q^2_{th} \sim 10
GeV^2$ and the $Q^2_{th}$ value may be larger or smaller to $Q^2_c$
one. Then, either the solution in the r.h.s. of
Eqs. (\ref{20},\ref{22},\ref{23})
before the critical point $Q^2_c$ and the one for $Q^2 > Q^2_c$ contain
the threshold transition, where the values of all variables are
changed from
ones at $f=3$ to ones at $f=4$. The $\alpha(Q^2)$ is smooth because
$\Lambda^{f=3}_{\overline{MS}} \to  \Lambda^{f=4}_{\overline{MS}}$
(see also the recent experimental test of the flavour independence of
strong interactions into \cite{17a}).

For simplicity here we suppose that $Q^2_{th} = Q^2_c$ and all changes
initiated by threshold are done authomatically: the first (at $Q^2
\leq Q^2_c$) solutions contain $f=3$ and second (at $Q^2 > Q^2_c$)
ones have $f=4$,
respectively. For the ``$-$'' component we should use $Q^2_{th}=Q^2_c$,
too.

Note only that the Pomeron intercept $\alpha_p = 1~-~(d_+ \tilde s +
\hat d^q_{++}p)$ increases at $Q^2=Q^2_{th}$, because
that agrees
\[
\alpha_p ~-~1 ~=~
\left\{
\begin{array}{ll}
\frac{4}{3} \tilde s(Q^2_{th},Q^2_0)~-~ \frac{2800}{81} p(Q^2_{th},Q^2_0)
 , & \mbox{ if } Q^2 \leq Q^2_c \\
1.44 \tilde s(Q^2_{th},Q^2_0)~-~ 38.11 p(Q^2_{th},Q^2_0)
,& \mbox{ if } Q^2>Q^2_c
\end{array} \right.
\]

with results \cite{18a} obtained in
the framework of dual parton model. The difference
$$ \bigtriangleup \alpha_p ~=~ 0.11 \tilde s(Q^2_{th},Q^2_0) - 3.55
p(Q^2_{th},Q^2_0) $$
dependes from the values of $Q^2_{th}$ and $Q^2_0$.
 For $Q^2_{th}=10GeV^2$ and $Q^2_0=1GeV^2$ it is very small:
$$ \bigtriangleup \alpha_p ~=~ 0.012 $$

\section{Discussion} \indent

Let us summarize the obtained results. We have got the DGLAP equation
``solution'' having the Regge form (\ref{1}) for the two cases: at small
$Q^2$ ($Q^2 \sim 1GeV^2$), where SF and PD have the flat behaviour at
small $x$, and at large $Q^2$, where SF $F_2(x,Q^2)$ fastly increases
when $x \to 0$. The behaviour in the flat case is nonstable with the
perturbative viewpoint because it leads to the production of the
supercritical value of Pomeron intercept at larger $Q^2$ and the its
increase (like $4/3~ ln(\alpha (Q^2_0)/\alpha(Q^2)$ in LO) when the
 $Q^2$ value increases\footnote{The Pomeron intercept value increasing with
  $Q^2$ was obtained also in \cite{19a,20a}.}. The solution in the Lipatov
Pomeron case corresponds to the well-known results (see
\cite{12,14,17}) with $Q^2$-independent Pomeron intercept. The general
``solution'' should contains the smooth transition between these
pictures. Unfortunately, it is impossible to obtain it in the case of
the simple approximation (\ref{1}), because the r.h.s. of DGLAP
equation (\ref{7}) contains  both: $\sim x^{-\delta}$ and $\sim
Const$ terms. As a result, we used the two above ``solutions'' gluing
them at
some point $Q^2_c$.

Note that our ``solution'' is some generation (or an application) of
the solution of the DGLAP equations in the momentum space. The last
one has two: ''+'' and ``$-$'' components. Our
conclusions are related to the ``+'' component, which exhibits the basic
Regge asymptotic behaviour. The Pomeron intercept corresponding to ``$-$''
component, is $Q^2$-independent and this component is the
subasymptotical one at large $Q^2$. However, the magnitude
of the ``+'' is suppressed
like $1/ln(1/x)$ and $\alpha (Q^2_0)$, and the subasymptotical ``$-$''
component may be important. Indeed, it is observed experimentally (see
\cite{20.5,20}). Note, however, that the suppression $\sim
\alpha(Q^2_0)$ is really very slight if we choose a small value of
$Q^2_0$.

Our ``solution'' in the form of Eqs.(\ref{22})-(\ref{25}) is in  very
good agreement with the recent $MRS(G)$ fit \cite{19} and with the
results of \cite{17} at $Q^2=15GeV^2$. As it can be seen from
Eqs.(\ref{22}),(\ref{23}), in our formulae there is the dependence
on the PD behaviour
at large $x$. Following \cite{21a} we choose $\nu =5$ that
agrees in the
gluon case with the quark counting rule \cite{22a}.
This $\nu$ value is also close to  the values obtained by
$CCFR$ group \cite{23a} ($\nu = 4$) and in the last $MRS(G)$ analysis
\cite{19} ($\nu =6$). Note that this dependence is strongly reduced
for the gluon PD in the form
$$ f_g(x,Q^2_0)~=~A_g(\nu)(1-x)^{\nu}, $$
if we suppose that the proton's momentum is carried by the gluon, is
 $\nu$-independent. We used $A_g(5)=2.1$ and $F_2(x,Q^2_0)=0.3$ when
$x \to 0$.

For the quark PD the choice $\nu =3$ is more preferable, however the
use of  two different $\nu$ values complicates the analysis. Because
the quark contribution to the ``+'' component is not large, we put
$\nu =5$ to both: quark and gluon cases. Note also that the variable
$\nu (Q^2)$ has (see \cite{24a}) the $Q^2$-dependence determined by
the LO AD $\gamma^{(0)}_{NS}$. However this $Q^2$-dependence is
proportional to $s$ and it is not important in our analysis.

Starting from $Q^2_0=1GeV^2$ (by analogy with \cite{25a}) and from
$Q^2_0=2GeV^2$, and using two values of the QCD parameter $\Lambda$:
the more
standard one ($\Lambda^{f=4}_{\overline {MS}}$ = 200 $MeV$) and
 ($\Lambda^{f=4}_{\overline {MS}}$= 255 $MeV$) obtained in \cite{19},
we have the following values of the quark and gluon PD ``intercepts''
$\delta_a ~=~-(\hat d_+ \tilde s +\hat d^a_{++}p)$ (here
$\Lambda^{f=4}_{\overline {MS}}$ is marked as $\Lambda$):

if $Q^2_0$ = 1 $GeV^2$
\begin{center}
\begin{tabular}{|l||l|l|l|l|}         \hline
$Q^2$ & $\delta_q(Q^2)$  & $ \delta_g(Q^2)$ & $\delta_q(Q^2)$  &
$\delta_g(Q^2)$ \\
    &$\Lambda =200MeV$ &$\Lambda =200MeV$ &$\Lambda =255MeV$ &$\Lambda
    =255MeV$ \\   \hline
4   &   0.191        &   0.389        &   0.165        &   0.447   \\ \hline
10  &   0.318        &   0.583        &   0.295        &   0.659   \\ \hline
15  &   0.367        &   0.652        &   0.345        &   0.734   \\ \hline
\end{tabular}
\end{center}

if $Q^2_0$ = 2 $GeV^2$
\begin{center}
\begin{tabular}{|l||l|l|l|l|}         \hline
$Q^2$ & $\delta_q(Q^2)$  & $ \delta_g(Q^2)$ & $\delta_q(Q^2)$  &
$\delta_g(Q^2)$ \\
    &$\Lambda =200MeV$ &$\Lambda =200MeV$ &$\Lambda =255MeV$ &$\Lambda
    =255MeV$ \\   \hline
4   &   0.099        &   0.175        &   0.097        &   0.198   \\ \hline
10  &   0.226        &   0.368        &   0.227        &   0.410   \\ \hline
15  &   0.275        &   0.438        &   0.278        &   0.486   \\ \hline
\end{tabular}
\end{center}

These values of $\delta_a $ are above at 4 $GeV^2$ those from
\cite{19}. Because we have the second (subasymptotical) part, the our
effective ``intercepts'' have smaller values.

Note that as input we can use the Eq.(\ref{1}) with $\delta \equiv
\varepsilon = 0.08$, which corresponds to Donnachie-Landshoff value
\cite{26a} of the Pomeron intercept. For this purpose we should
represent the value $1/ \tilde \varepsilon $ (see Eq.(\ref{3})) as the
series (here $\Psi^{'}(\nu)\equiv \frac{d}{d\nu } \Psi(\nu)$):
 \begin{eqnarray}  \z
 \frac{1}{\tilde \varepsilon}~\equiv~\frac{1}{\varepsilon}
\biggl[ 1 - \frac{\Gamma(\nu +1)\Gamma(1- \varepsilon)}
{\Gamma(\nu +1 - \varepsilon)}x^{\varepsilon} \biggr] ~=~ ln \frac{1}{x} -
\Bigl[ \Psi(\nu +1) - \Psi(1) \Bigr] +    \nonumber \\ \z
\frac{\varepsilon}{2} \biggl[ {\biggl(  ln \frac{1}{x} -
\Bigl[ \Psi(\nu +1) - \Psi(1) \Bigr] \biggr)}^2 -
\Bigl[ \Psi^{'}(\nu +1) - \Psi^{'}(1) \Bigr]  \biggr] +
O(\varepsilon ^2)
 \label{26}
 \end{eqnarray}
and save only the first term in the r.h.s. This is possible if the
second term in the r.h.s. of Eq.(\ref{26}) is negligibly small, i.e.
$$ \frac{\varepsilon}{2} \biggl[ {\biggl(  ln \frac{1}{x} -
\Bigl[ \Psi(\nu +1) - \Psi(1) \Bigr] \biggr)}^2 -
\Bigl[ \Psi^{'}(\nu +1) - \Psi^{'}(1) \Bigr]  \biggr] \ll  ln \frac{1}{x} -
\Bigl[ \Psi(\nu +1) - \Psi(1) \Bigr], $$
or for $\nu =5$
$$1.5 \cdot 10^{-12} \ll x \ll 9.6 \cdot 10^{-2}$$

This $x$ range corresponds to the range considered here. The right
boundary slowly depends on the $\delta$ value: it is determined by the
large $x$ SF behaviour, i.e. by the $\nu$ value. The left boundary
strongly increases with increasing of the $\delta$ value, that leads
to the impossibility to apply the expansion (\ref{26}) for the large
$\delta$ values.

In the case $\delta = \varepsilon$ all our conclusions are not changed
 except the relation $\alpha_p = 1+ \delta (Q^2)$,
which transforms to the form $\alpha_p = 1+ \varepsilon + \delta
(Q^2),$
i.e. the value of the Pomeron intercept slightly increases.

In support of our analysis we represent the Fig.3 of paper \cite{5} and add
it by our results for $\delta_q (Q^2)$ of above table at $Q^2_0 = 1
{}~GeV^2$.
The dashed-dotted curve represents the intercept of the Pomeron
trajectory $\alpha_p(Q^2)$
which was obtained in \cite{19a} as the result of a fit of
experimental data. This analysis was done in the framework of Regge-type
behaviour of DIS SF, that corresponds to the start point of our
consideration, too.
 The authors of \cite{19a} obtained very strong
$Q^2$-dependence of the Pomeron intercept in the region of
$1~GeV^2<Q^2<10~GeV^2$
and approximate $Q^2$-independent its values $\alpha_p \simeq 1.05$
and  $\alpha_p
\simeq 1.4$ at $Q^2<1~GeV^2$ and  $Q^2>10~GeV^2$, respectively. The solid
curves represent our values of $Q^2$-dependence of Pomeron intercept
 \begin{eqnarray}
\alpha_p(Q^2)  ~=~
\left\{
\begin{array}{ll}
1.05 + \delta_q(Q^2)
 , & \mbox{ if } Q^2 \leq Q^2_c \\
1.05 + \delta_q(Q^2_c)
,& \mbox{ if } Q^2>Q^2_c,
\end{array} \right.
 \label{aa}
 \end{eqnarray}
where we choose $\alpha_p(Q^2_0)= 1.05$ and $Q^2_c = 15 GeV^2$. As can
be seen in the Figure both the results are in
 very good agreement.

As a conclusion, we note that BFKL equation (and thus the value of
Lipatov Pomeron intercept) was obtained in \cite{19.5} in the
framework of perturbative QCD. The large-$Q^2$ $HERA$ experimental
data are in the good agreement with Lipatov's trajectory and thus with
perturbative QCD. The small $Q^2$ data agrees with the standard
Pomeron intercept $\alpha_p=1$ or with Donnachie-Landshoff picture:
$\alpha_p=1.08$. Perhaps, this range requires already the knowledge of
nonperturbative QCD dynamics and  perturbative solutions (including
BFKL one) should be not applied here directly and should be corrected by some
 nonperturbative contributions (see \cite{27a}).

\section{Conclusions} \indent

Thus, in our analysis Eq.(\ref{1}) can be considered as the
nonperturbative
(Regge-type) input at $Q^2_0 \sim 1GeV^2$. Above $Q^2_0$ the PD
behaviour obeys DGLAP equations. Under the action of perturbative QCD
the Pomeron splits into two components. The intercept of the ``$-$''
component is $Q^2$ independent and this component is the
subasymptotical one at larger $Q^2$ values. The Pomeron corresponding
to the ``+''
component  moves to the supercritical
regime and tends to its perturbative value. Above some $Q^2_c$, where
its perturbative value is already  attained, the Pomeron intercept keeps
a constant value. Our analysis supports the idea \cite{19a,20a} about
the one effective Pomeron having a $Q^2$-dependent intercept, however
the character of the $Q^2$-dependence is different.

 The application of this approach to analyse small
$x$ data and taking into account a resummation of all $\alpha_s $-orders
of perturbative QCD (by analogy with \cite{14a})
invite further investigation and will be considered in future.


\vspace{1cm}
\hspace{1cm} \Large{} {\bf Acknowledgements}    \vspace{0.5cm}

\normalsize{}

Author is grateful to Professors Aurenche, Capella, Donnachie, Fadin,
Kwiecinski and Lipatov  for discussions.

\section{Appendix} \indent
\def\theequation{A\arabic{equation}}
\setcounter{equation}0

  Here we present the illustration of the method to replace the
convolution
  by simple product at small$~x$.  More detailed analysis can
be found in \cite{13}.

  {\bf 1.} Consider the basic integral
  $$J_{\delta}(a,x)=x^a*\varphi(x) \equiv
  \int_{x}^{1}\frac{{\rm d}y}{y}~y^a~ \varphi(\frac{x}{y}),$$
where $\varphi(x) = Ax^{-\delta}(1-x)^{\nu}
\equiv x^{-\delta}\tilde\varphi(x)$.
Expanding $\tilde\varphi(x)$ near
$\tilde\varphi(0)$ ,
 we have
 \begin{eqnarray} J_{\delta}(a,x) &=&
 x^{-\delta}\int_{x}^{1}{\rm d}y~ y^{a+\delta-1} \left[
 \tilde\varphi(0)+\frac{x}{y}~ \tilde\varphi^{(1)}(0)+ \ldots
  +\frac{1}{k!} \left(\frac{x}{y}\right)^k \tilde\varphi^{(k)}(0)+ \ldots
 \right] \nonumber  \\
&=& x^{-\delta}
 \left[ \frac{1}{a+\delta}~ \tilde\varphi(0)+
 O(x) \right]  \label{A1}  \\
&-& x^a
 \left[ \frac{1}{a+\delta}~ \tilde\varphi(0)+
  \frac{1}{a+\delta-1}~ \tilde\varphi^{(1)}(0)+ \ldots
  +\frac{1}{k!}~ \frac{1}{a+\delta-k}~ \tilde\varphi^{(k)}(0)+ \ldots
 \right] \nonumber
\end{eqnarray}
The second term on the r.h.s. of eq.(\ref{A1}) can be summed, and
$J_{\delta}(a,x)$   has the following form
$$J_{\delta}(a,x)=x^{-\delta}
\left[\frac{1}{a+\delta}~ \tilde\varphi(0)+
O(x) \right]~ + ~
x^a~ \frac{\Gamma(-(a+\delta))\Gamma(1+\nu)}{\Gamma(1+\nu-a-\delta)}~
\tilde\varphi(0)$$
Consider two important cases:

{\bf a)} $a \geq 1$
$$J_{\delta}(a,x)=x^{-\delta}\frac{1}{a+\delta}~
\tilde\varphi(
x) +
O(x^{2- \delta})$$

{\bf b)} $a=0$
$$J_{\delta}(0,x)=x^{-\delta}
\left[\frac{1}{\delta}~ \tilde\varphi(0)+
O(x) \right]~ + ~
\frac{\Gamma(-\delta)\Gamma(1+\nu)}{\Gamma(1+\nu-\delta)}~
\tilde\varphi(0)=
x^{-\delta}~\frac{1}{\tilde\delta}
{}~\tilde\varphi(
x) +
O(x^{1- \delta})$$
where
$$  \frac{1}{\tilde \delta}~=~\frac{1}{\delta}
\Bigl( 1 -  \frac{\Gamma(\nu +1)\Gamma(1-\delta)}{\Gamma(\nu
  +1-\delta)}x^{\delta} \Bigr) $$

  {\bf 2.} Consider the integral
$$I_{\delta}(x)=\hat K(x)*\varphi(x) \equiv
 \int_{x}^{1}\frac{{\rm d}y}{y}~\hat K(y)~ \varphi(\frac{x}{y})$$
and define the moments of the kernel $\hat K(y)$ in the following form
$$K_n= \int_{0}^{1}{\rm d}y~y^{n-2}\hat K(y)$$

In analogy with subsection {\bf 1} we have
 \begin{eqnarray} I_{\delta}(x) &=&
 x^{-\delta}\int_{x}^{1}{\rm d}y~ y^{\delta-1}~\hat K(y)~ \left[
 \tilde\varphi(0)+\frac{x}{y}~ \tilde\varphi^{(1)}(0)+ \ldots
  +\frac{1}{k!} \left(\frac{x}{y}\right)^k \tilde\varphi^{(k)}(0)+ \ldots
 \right] \nonumber  \\
&=& x^{-\delta}
 \left[ K_{1+\delta}~ \tilde\varphi(0)+
 O(x) \right]  \label{A2}  \\
&-& x^a
 \left[ N_{1+\delta}(x)~ \tilde\varphi(0)+
  N_{ \delta}(x)~ \tilde\varphi^{(1)}(0)+ \ldots
  +\frac{1}{k!}~ N_{1+\delta-k}(x)~ \tilde\varphi^{(k)}(0)+ \ldots
 \right], \nonumber
\end{eqnarray}
where
$$N_{\eta}(x)= \int_{0}^{1}{\rm d}y~y^{\eta-2}\hat K(xy)$$

The case $K_{1+ \delta} = 1/(a + \delta)$ corresponds to $\hat
K(y)=y^a$
and has
been already considered in subsection { \bf 1}. In the more general cases
(for example, $K_{1+ \delta} = \Psi( \delta) + \gamma$) we can represent
the "moment" $K_{1 + \delta}$ as the combination of singular and regular
(for $ \delta \rightarrow 0 $) parts, i.e. $K_{1+ \delta} = -1/\delta +
\Psi(1+ \delta) + \gamma$. For the singular term the analysis from
subsection {\bf 1} may be repeated. As the regular part can be represented by
series of the sort
$ \sum_{m=1} 1/(a + \delta + m)$, then any additional contributions
from term $N_{1 + \delta}(x) \varphi (0)$ to any term of the series, are not
necessary.

So, for the initial integral at small $x$ we get the simple equation:

$$I_{\delta}(x)=x^{-\delta}~ \tilde K_{1+\delta}~
\tilde\varphi(
x) +
O(x^{1- \delta})$$
where $\tilde K_{1+\delta}$
 coincides with
$K_{1+\delta}$
after the replacement $1/\delta\rightarrow 1/\tilde\delta$.


\newpage

{ \bf Figure 1.} The intercept of the Pomeron trajectory
 $\alpha_p(Q^2)$ (dashed-dotted line) as obtained from the ALLM parametrization
(see \cite{5,19a}). The dotted lines show the uncertanty of the fit.
The solid curves represent the values of  $\alpha_p(Q^2)$ from Eq.(\ref{aa}).

\end{document}